\documentclass[dvips]{article}
\usepackage{icrctc07}

\newcommand{\gep}{G_{Ep}}
\newcommand{\gmp}{G_{Mp}}
\newcommand{\gmn}{G_{Mn}}
\newcommand{\gmpmu}{G_{Mp}/\mu_{p}}

\newcommand{\gen}{G_{En}}
\newcommand{\gmnmu}{G_{Mn}/\mu_{n}}

\title{Modeling Atmospheric Neutrino Interactions: Duality Constrained Parameterization of Vector and Axial
Nucleon Form Factors}
\shorttitle{Vector and Axial Form Factors}
\authors{ A. Bodek, S. Avvakumov, R. Bradford, and  H. Budd}
\shortauthors{A. Bodek et al}
\afiliations{Department of Physics and Astronomy,
University of Rochester, Rochester, NY  14627-0171}
\email{Presented by Arie Bodek, Email: bodek@pas.rochester.edu}

\abstract{
We present new
parameterizations of vector and axial nucleon form factors.
We maintain an excellent descriptions of the
form factors at low momentum transfers, where the spatial
structure of the nucleon is important, and
use the Nachtman scaling variable $\xi$
to relate elastic and inelastic form factors
and impose quark-hadron duality constraints at high momentum
transfers where the quark structure dominates. 
We use the new vector form factors to re-extract
updated values of the axial form factor from neutrino
experiments on deuterium. We obtain an updated world
average value from neutrino
and pion electroproduction experiments of $M_{A}$ = $1.0144 \pm 0.0136~GeV/c^2$.
Our parameterizations are useful in modeling atmospheric neutrino interactions
(e.g. for neutrino oscillations experiments). }

\begin{document}
\maketitle
%Begin the section.

% \section{Introduction}

At low $Q^2$, a reasonable  description of the proton and neutron
elastic form factors is given by the dipole approximation.
%---------------------EQUATION----------------------------------
    \begin{eqnarray}
G_D^{V,A}(Q^2) =
{\frac{C^{V,A}}{\left(1+\frac{Q^2}{M_{V,A}^2}\right)^2}}, \nonumber 
 \end{eqnarray}
%---------------------EQUATION----------------------------------
where $C^{V,A}$= (1,$g_{A}$), $g_{A}$ = -1.267, $M_{V}^2$ = 0.71 $(GeV/c^2)^2$,
  and $M_{A}$ = 1.015 $GeV/c^2$.
  
Here we present
parameterizations that simultaneously satisfy constraints at low $Q^2$ where the
spatial structure of the nucleon is important, as well as at high
$Q^2$ where the quark structure is important. Our new quark-hadron
duality based  parameterization (``BBBA07'') includes:
(1) Improved functional form that uses Nachtman scaling variable $\xi$
to relate elastic and inelastic vector and axial nucleon form factors;
(2) Yield the same values as $fits$ by
Arrington and Sick \cite{kelly} for $Q^{2}< 0.64
(GeV/c)^{2}$,
while satisfying 
quark-hadron duality constraints at high-$Q^2$.

We use our new ``BBBA07''vector form factors to re-extract updated values of the
axial form factor from a re-analysis of previous neutrino scattering data 
on deuterium and present a new parameterization for the axial form
factor within the framework of quark-hadron duality. 
%
%--------------------------------FIGURE---1---BEGIN----------
\begin{figure*}[th]
 \begin{center}
\includegraphics [width=0.96\textwidth]{{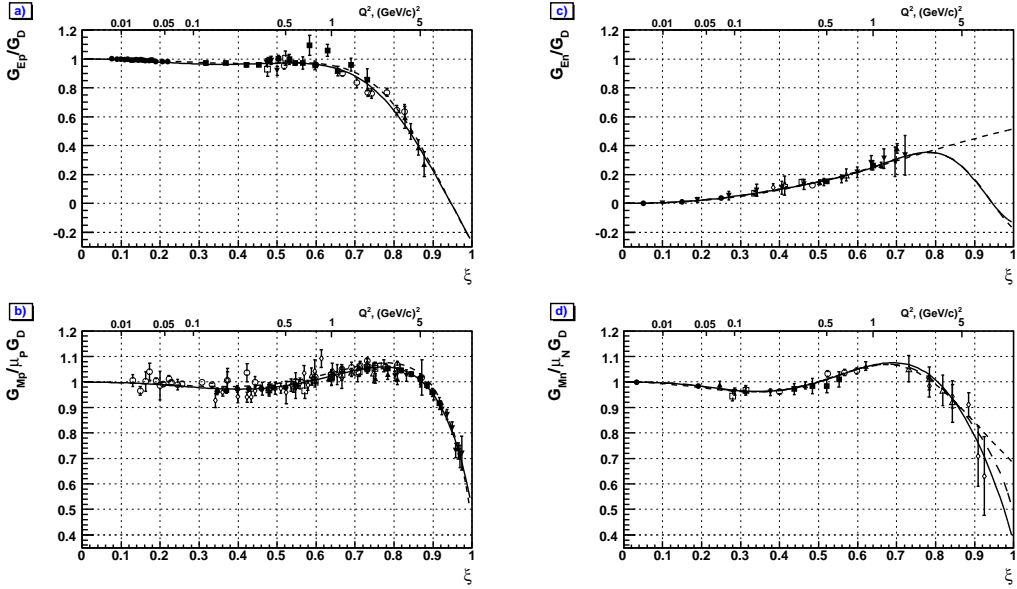}}
    \end{center}
   \caption[Ratios to $G_D$]{Ratios of $\gep$ (a), $\gmpmu$ (b),
   $\gen$ (c) and
$\gmnmu$ (d) to $G_D^{V}$.   
The short-dashed line in each plot is
the old Kelly parameterizations (old Galster for $\gen$).  The solid line is our new BBBA07
parameterization for $\frac{d}{u}=0.0$, and the long-dashed line is
 BBBA07 for $\frac{d}{u}=0.2$.
}
   \label{subfig1}
\end{figure*}
%-----------------------------FIGURE 1-----END-------------------
%
For vector form factors our $fit$ functions are $A_N(\xi)$ (i.e. 
$A_{Ep}(\xi)$, $A_{Mp}(\xi)$, $A_{En}(\xi)$, $A_{Mn}(\xi)$)
multiplied by an updated Kelly  \cite{kelly} type parameterization
 of one of the proton form factors.
 %---------------------EQUATION----------------------------------
 \begin{eqnarray}
A_N (\xi) &=& \sum_{j=1}^{n} P_j (\xi) \nonumber  \\
  P_j (\xi) &=& p_j\prod_{k=1, k \ne j}^{n} \frac{\xi - \xi_k}{\xi_j -
\xi_k}. \nonumber\\
  G^{Kelly}(Q^2) &=& \frac{\sum_{k=0}^{m}a_k \tau^k}{1 +
\sum_{k=1}^{m+2}b_k \tau^k}, \nonumber
\end{eqnarray}
%---------------------EQUATION----------------------------------
where $a_{0}=1$, $m=1$, and $\tau = Q^2/4M_{N}^2$.
($M_N$ is proton, neutron, or average nucleon mass for proton,
neutron, and axial form factors, respectively).
The
datasets used by Kelly to $fit$ $\gep$ and $\gmpmu$ ($\mu_p = 2.7928$,
$\mu_n = - 1.913$) are described in
\cite{kelly}.  Our parameterization employs the as-published Kelly
parameterization to $G^{Kelly}_{Ep}$ and an updated set of parameters
for $G^{Kelly}_{MP}(Q^2)$ that includes the
recent BLAST\cite{crawford} results. The
parameters used for  $G^{Kelly}(Q^2)$ are given in
ref. \cite{bodek}.
Each $P_j$ is a LaGrange polynomial in the Nachtman variable,
$\xi=\frac{2}{(1+\sqrt{1+1/\tau})}$ 
The
  $\xi_j$ are equidistant ``nodes'' on an interval $[0,1]$ and
$p_j$ are the
$fit$  parameters that have the property $A_N (\xi_j) = p_j$.
The seven $p_j$ parameters are
at $\xi_j$=0, 1/6, 1/3, 1/2, 2/3, 5/6, and 1.0.
%---------------------

In the $fitting$  procedure described below, the parameters of $A_N(\xi)$ are
constrained
to give the same vector form factors as the recent low $Q^2$
$fit$ of Arrington and Sick \cite{kelly} for $Q^{2}< 0.64
(GeV/c)^{2}$ (as that analysis includes coulombs corrections which 
modify $G_{Ep}$,two photon exchange corrections which modify
$G_{Mp}$ and $G_{Mn}$). Since the published form 
factor data do not have these corrections, this constraint is 
implemented by including additional $''fake''$ data points for $Q^{2}<
0.64 (GeV/c)^{2}$.
Our fits to the  form factors are:
%---------------------EQUATION----------------------------------
\begin{eqnarray}
      {G_{Mp}(Q^2)} & = & {\mu_{p}}{ A_{Mp}(\xi)} \times  
      {G^{Kelly}_{Mp}(Q^2)} \nonumber \\
 {G_{Ep}(Q^2)}  &=&  A_{Ep}(\xi)\times {G^{Kelly}_{Ep}(Q^2)} \nonumber \\
   {G_{En}(Q^2)}  &=&  A^{a,b}_{En}(\xi)\times {G_{Ep}(Q^2)} \times
   \left( {\frac{a\tau}{1+b\tau}} \right) \nonumber \\
   {G_{Mn}(Q^2)} &=&  A^{a,b}_{Mn}(\xi)
    \times  {G_{Mp}(Q^2)} {\mu_{n}} /{\mu_{p}}, \nonumber 
  \end{eqnarray}
%---------------------EQUATION----------------------------------
  where we use 
 our updated parameters in the Kelly parameterization of $\gmp$.
 For $\gen$ the parameters a=1.7 and b=3.3 are the same as
 in the Galster\cite{kelly}  parametrization and ensure that 
 $d\gen/dQ^2$ at for $Q^{2}=0$ is in agreement with measurements.
  The values  $A(\xi)$=$p_1$ at $\xi_1$=0  ($Q^2 =0$) for
   $\gmp$,  $\gep$,   $\gen$,   $\gmn$ are set to  
  to 1.0. The value  $A(\xi)$=$p_7$ at $\xi_j$=1 
  ($Q^2 \rightarrow \infty$) for $\gmp$ and $\gep$ is set to 1.0.
   The value  $A(\xi)$=$p_j$ at $\xi_j$=1 
for   $\gmn$ and  $\gen$ are $fixed$ by  constraints from quark-hadron
duality.  
Quark-hadron duality implies that the ratio of neutron and
proton magnetic form factors should be the same as the ratio of the
corresponding inelastic structure functions $ \frac{F_{2n}}{F_{2p}} $
in the $\xi$=1 limit. (Here $F_2= \xi \sum_{i}e_i^2 q_i \left ( \xi \right)$)
%---------------------EQUATION----------------------------------  
\begin{eqnarray}
\frac{G_{Mn}^2}{G_{Mp}^2}= \frac{F_{2n}}{F_{2p}} = {
\frac{1+4\frac{d}{u}}{4+\frac{d}{u}} = \left(
\frac{\mu_{n}^{2}}{\mu_{p}^{2}}\right)
A_{Mn}^{2}(\xi=1) } \nonumber
 \end {eqnarray}
%---------------------EQUATION----------------------------------
 %------------------------------------------FIGURE--2-  BEGIN
\begin{figure}
    \begin{center}
\includegraphics [width=0.48\textwidth]{{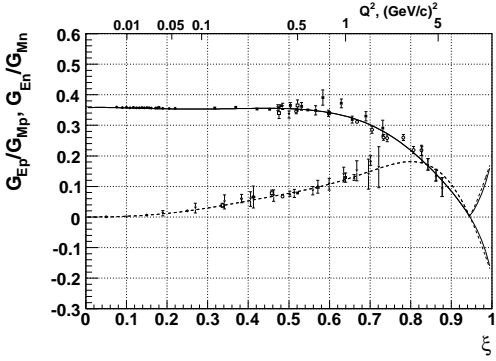}}
\end{center}
\caption[$G_{En}$]{The constraint used in $fit$ting $G_{En}$
stipulates that $\gen^2/\gmn^2=\gep^2/\gmp^2$ at high $\xi$.  
The solid line is $ \frac{G_{Ep}}{|G_{Mp}|}$ and $\frac{|G_{Ep}|}{|G_{Mp}|}$, 
and the short-dashed
line is $\frac{G_{En}}{|G_{Mn}|}$  and $\frac{|G_{En}|}{|G_{Mn}|}$.
}
\label{subfig2}
\end{figure} 
%---------------------------------------------FIGURE 2 END------

 We ran $fits$
with two different values of $\frac{d}{u}$ at the $\xi$=1 limit:
$\frac{d}{u}$ =0 and 0.2 (corresponding
to $\frac{F_{2n}}{F_{2p}}$  = 0.25 and 0.4286). 
The $fit$ utilizing $\frac{d}{u}=0$ is $A^{25}_{Mn}$, and
 the $fit$ utilizing $\frac{d}{u}=0.2$ is $A^{43}_{Mn}$.  
The value  $A(\xi)$=$p_j$ at $\xi_j$=1 
for   $\gen$ is set by another duality-motivated
constraint.  $R$ is $defined$ as the ratio of deep-inelastic 
longitudinal and transverse structure functions.
In the elastic limit: 
$$R_n( x=1; Q^2) = (4M_{N}^2/{Q^2})\times (G^2_{En}/G^2_{Mn}).$$
For inelastic scattering,
as $Q^2 \rightarrow \infty$, $R_n=R_p$. If we assume quark-hadron
duality, the same is for the elastic form factors
at $\xi$=1 ( $Q^2 \rightarrow \infty$)
%---------------------EQUATION----------------------------------
\begin{eqnarray}
      {G^2_{En}}/{G^2_{Mn}}= {G^2_{Ep}}/{G^2_{Mp}} \nonumber
\end{eqnarray}
%---------------------EQUATION----------------------------------
The new form factors $\gep$, $\gmpmu$, $\gmnmu$, and $\gen$ are plotted
in Figure 1 as ratios to the dipole form $G_D^{V}$. 
as ratios to the dipole form factor, $G_D^{V}$.
 $A_N(\xi)$ is not needed for  $\gmp$ as 
it is very close to 1.0.  For $\gep$ it yields a correction
of $1\%$  at low $Q^2$ (because it is required to agree with the
$fits$ 
of Arrington and Sick\cite{kelly} which include two photon exchange and Coulomb
corrections. For $\gen$ and $\gmn$  it is used to impose 
 quark-hadron duality asymptotic  constraints.
Figure \ref{subfig2} shows plots of $\frac{G_{En}}{|G_{Mn}|}$ and
$\frac{G_{Ep}}{|G_{Mp}|}$ (for the  $\frac{d}{u}$ = 0 at $\xi$ = 1
case). The long-dashed line is a simple quark-hadron
duality model \cite{qduality}.

Using the updated vector form factors, we perform a complete
reanalysis of  published neutrino quasielastic \cite{neutrinoD2} data on
deuterium. (Because of uncertain nuclear corrections, neutrino data on 
heavier nuclear targets are not used.) 
We extract new values of $M_A$, and updated values of $F_A(Q^2)$.
The average of the corrected neutrino measurements is $M_{A}$  =
$1.0156 \pm0.0278$.  This is to be compared to the average value
of $1.014 \pm0.016$ extracted from pion electroproduction experiments\cite{pion}
after corrections for hadronic effects. The average of the two average
values is $1.0144 \pm0.0136$.
%----------------------------------BEGIN TABLE MA
\begin{table} 
\begin{center}
\begin{tabular}{|l|l|l|l|}
$Experiment $ & $M_A$& $\Delta M_A$	\\
& (published) &  new--old    \\
\hline 
    $Miller-D-ANL_{82,77,73}$      & 1.00 $\pm$ 0.05      & $-0.030$ \\
     $Baker-D-BNL_{81}$        & 1.07 $\pm$ 0.06      & $-0.028$  \\
 $Kitagaki-D-FNAL_{83}$  & 1.05$_{-0.16}^{+0.12}$& $-0.025$    \\ 
 $Kitagaki-D-BNL_{90}$  & 1.070$_{-0.045}^{+0.040}$ & $-0.036$   \\ 
\end{tabular}
\end{center}
\caption{$M_A$ $GeV/c^2$ values published
by neutrino-deuterium experiments 
and updated corrections  $\Delta M_A$ when re-extracted with updated
vector form factors. 
 }
\label{MA_values}
\end{table}
%----------------------------------END TEBLE MA
%-------------------------BEGIN FIGURE 2
\begin{figure*}
 \begin{center}
\includegraphics [width=0.96\textwidth]{{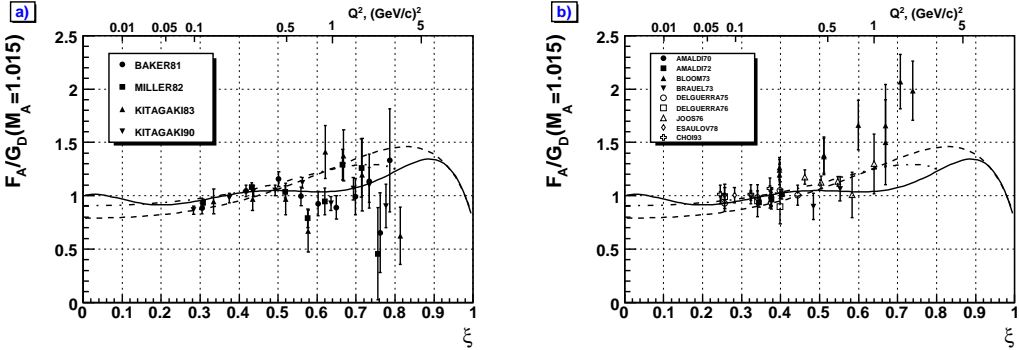}}
 \end{center}
 \caption[$F_{A}$:Axial Form factor ratios to $G_D^{A}$]{
 (a) $F_A(Q^2)$ re-extracted from neutrino-deuterium
 data divided by $G_D^{A}(Q^2)$ (with $M_{A} = 1.015$). (b)  
$F_A(Q^2)$ from pion electroproduction
  divided by $G_D^{A}(Q^2)$ (corrected for
  for hadronic effects\cite{pion}). 
  The solid line is our duality based
  $fit$. The short-dashed line is  $F_A(Q^2)_{A2=V2}$. 
  The dashed-dot line
  is a prediction from a constituent quark model.
}
\label{figaxial}  
\end{figure*}
 The new form factors $\gep$, $\gmpmu$, $\gmnmu$, and $\gen$ are plotted
in Figure 1.
%---------------------------------------------END FIGURE 3

  For deep-inelastic scattering, the vector and axial
  parts of the inelastic structure functions $W_{2}$ 
  are equal.  Local quark-hadron duality 
    at large $Q^2$ implies  that
  the axial and vector components of $W_{2}^{elastic}$
  are also equal, yielding: 
%---------------------EQUATION----------------------------------
  \begin{eqnarray}
{[F_A(Q^2)_{A2=V2}]^{2}} &=& \frac {(G_E^{V})^{2}(Q^2)+\tau 
(G_M^V(Q^2))^{2}}{(1 + \tau )}, \nonumber \\
 \end {eqnarray}
%---------------------EQUATION----------------------------------
 where $ G_E^V(Q^2)=G_{Ep}(Q^2)-G_{En}(Q^2) $ and  
  $G_M^{V}(Q^2) = G_{Mp}(Q^2)-G_{Mn}(Q^2)$.
 We do a duality based $fit$ to the updated values of 
 the axial form factor $F_A(Q^2)$, including pion electroproduction
 data.
 Here the $fit$ function is a sum of
LaGrange polynomials, $A^{a}_{FA}$, multiplied by $G_D^{A}(Q^2)$
(with $M_{A} = 1.015$).
%---------------------EQUATION----------------------------------
\begin{eqnarray}
F_A (Q^2)=A^{a}_{FA} (\xi) \times G_D^{A}(Q^2). \nonumber
\end{eqnarray}
%--------------------EQUATIOM----------------------------------
We impose the constraint  $A^{a}_{FA} (\xi_1=0) = p_1 = 1.0$. We also
constrain the $fit$
by requiring
that $A^{a}_{FA}(\xi)$ yield  
$F_A(Q^2)=F_A(Q^2)_{A2=V2}$ for $\xi$ $> 0.9$  ($Q^{2} >7.2 (GeV/c)^2$). 
Figure 3(a) shows 
 values  of $F_A(Q^2)$  extracted from neutrino-deuterium experiments 
  divided by $G_D^{A}(Q^2)$, with $M_{A} = 1.015$.
Figure 3(b)  shows 
 values of $F_A(Q^2)$  extracted from pion electroproduction
 experiments 
  divided by $G_D^{A}(Q^2)$. These pion electroproduction
  values can be directly compared to the neutrino results because they
  are  multiplied  by a factor
  $F_{A}(Q^{2},M_{A}=1.014)$/$F_{A}(Q^{2},M_{A}=1.069)$  
  to correct  for $\Delta M_{A} = 0.055$ originating from 
   hadronic effects\cite{pion}.  
  The solid line is our duality based
  $fit$. The short-dashed line is  $F_A(Q^2)_{A2=V2}$. 
  The long-dashed line is $F_A(Q^2)_{A1=V1}$. The dashed-dot line
  is a prediction from a constituent quark model\cite{quark}.
Our new
parameterizations of vector and axial nucleon form factors
use quark-hadron duality constraints at high momentum
transfers and maintain a very good descriptions of the
form factors at low momentum transfers.
These parameterizations
are useful in modeling neutrino interactions 
(e.g. for neutrino oscillations experiments). 
Our predictions
for $\gen (Q^{2}) $ and $F_{A}(Q^{2})$ at high $(Q^{2}$)
can be tested in upcomng electron
scattering and neutrino
 experiments
at Jefferson Laboratory and at Fermilab (MINERvA).
The $final$ parameters are given in Ref. \cite{bodek}.


\begin{thebibliography}{20}
    
\bibitem{bodek} $www.pas.rochester.edu/~bodek/FF/$ includes computer
code for BBA2207 form factors.
A. Bodek \textit{et al} to be submitted to
Phys. Rev. Lett. 2007.



\bibitem{crawford}C.B. Crawford \textit{et al}, Phys. Rev. Lett 98,
052301 (2007).

\bibitem{kelly}J.J. Kelly, Phys. Rev. C 70, 068202 (2004);
S. Galster \textit{et al}, Nucl. Phys. B32, 221 (1971);
J. Arrington  I.Sick (Basel U.)
Submitted to Phys.Rev.C, nucl-th/0612079. 

\bibitem{qduality} e.g. if  $R_{elastic} =R_{inelastic}$ and $R_{inelastic}$
is dominated by target mass effects.
A. Bodek and U. K. Yang, Nucl.Phys.Proc.Suppl.139:113-118.2005.
S. Kretzer and H. M.  Reno, Phys. Rev. D69,034002 (2004).
;H. Georgi and H. D. Politzer, Phys. Rev.
D14, 1829 (1976); R. Barbieri et al., Phys.
Lett. B64, 171 (1976), and Nucl. Phys.B117, 50 (1976)

\bibitem{pion}V. Bernard, L. Elouadrhiri , U. Meissner, J. Phys. G28, R1
(2002). 

\bibitem{neutrinoD2}
N.J.~Baker {\em et al.}, Phys. Rev. D23 (1981)
2499;  K.L.~Miller {\em et al.}, Phys. Rev. D26 (1982) 537;
T.~Kitagaki {\em et al.}, Phys. Rev. D28 (1983)
436; T.~Kitagaki {\em et al.}, Phys. Rev. D42 (1990)  1331



\bibitem{quark} R. F., Wegenbrunn, \textit{et al}, hep-ph/0212190.


\end{thebibliography}
\end{document}